\documentclass[universe,article,accept,moreauthors,pdftex]{Definitions/mdpi}

\firstpage{1} 
\pubvolume{xx}
\issuenum{1}
\articlenumber{5}
\pubyear{2020}
\copyrightyear{2020}
\history{Received: date; Accepted: date; Published: date}

\pdfoutput=1

\usepackage{amssymb}
\usepackage{bm}
\usepackage{xcolor}
\usepackage{textgreek}

\Title{Relativistic effects in orbital motion of the S-stars at the Galactic Center}

\Author{Rustam Gainutdinov$^{1,2,*}$\orcidA{} and Yurij Baryshev$^{2}$\orcidB{}
}

\AuthorNames{Rustam Gainutdinov and Yurij Baryshev}

\address{%
$^{1}$ \quad SPb Branch of Special Astrophysical Observatory of Russian Academy of Sciences, 65 Pulkovskoye Shosse, St Petersburg 196140, Russia;\\
$^{2}$ \quad Saint Petersburg State University, 7/9 Universitetskaya Nab., St Petersburg 199034, Russia.}

\corres{Correspondence: roustique.g@gmail.com}

\abstract{The Galactic Center star cluster, known as S-stars, is a perfect source of relativistic phenomena observations. The stars are located in the strong field of relativistic compact object Sgr A* and are moving with very high velocities at pericenters of their orbits. In this work we { consider motion of several S-stars by using}  the Parameterized Post-Newtonian (PPN) formalism of General Relativity (GR) and Post-Newtonian (PN) equations of motion of the { Feynman's quantum-field gravity theory, where the positive energy density of the gravity field can be measured via the relativistic pericenter shift. The PPN parameters \texorpdfstring{$\beta$}{\textbeta} and \texorpdfstring{$\gamma$}{\textgamma} are constrained using the S-stars data. The positive value of the \texorpdfstring{$T_g^{00}$}{Tg00} component of the gravity energy-momentum tensor is confirmed for condition of S-stars motion.}
}


\begin{document}
\section{Introduction}

The S-stars cluster in the Galactic Center { \citep{Ghez2008,Genzel2010,Gillessen2009,Gillessen_2017}} is a unique observable object to investigate. Once for the orbital period the stars pass their pericenters, where they reach high velocities ($\sim$ 0.01 { or even $\sim$0.1 of the speed of light for the most recent discovered stars \cite{Iorio2020}}) and get close to the relativistic compact object Sgr A*. Such conditions lead to manifestation of relativistic effects in proximity to a supermassive compact object (Super-Massive Black Hole). So we are able to explore them in the frame of Post-Newtonian formalism. { For recent overviews of GR see, e.g., \citet{Debono2016} and references therein.}
\par
The Post-Newtonian formalism is a tool which provides an opportunity to express the relativistic equations of gravity as the small-order deviations from the Newtonian theory. The Parametrised Post-Newtonian formalism can be implemented to various theories of gravity {\citep{will_2018}}. It is a version of the PN formalism that has a parameters that quantitatively express the difference of a theory of gravity and the Newtonian theory. In this work, the parameters of $\beta$ and $\gamma$, which appear in the expansion of the Schwarzschild solution \citep{MTW1973}, are taken into consideration. Even in the Solar system, physical effects can be detectable via the PPN formalism. The most {precise} constrain of the $\gamma$ parameter {(with the accuracy of $2\times 10^{-5}$)} was obtained using the data from the Cassini experiment \citep{Bertotti_2003}. The detected { effect was the the time delay, which is also known as} the classical fourth test of general relativity by \citet{Shapiro_1964}. {The PPN parameters $\gamma$ and $\beta$ were also measured with the EPM2017 ephemerides \citep{Iorio2019,Pitjevy2018}. It turns out that $\beta$ is at the $10^{-5}$ level as well, and that also $\gamma$ measurement is reaching the Cassini experiment level.  In particular, \citep{Park2017} delivers $\mathrm{err}_{\beta} = 3.9 \times 10^{-5}$ and \citep{Genova2018} yields $\mathrm{err}_{\beta} =1.8 \times 10^{-5}$. \citet{Iorio2019}, relying uponrecent EPM data, suggests a $10^{-6}$ level.} In the modern Solar System experiments, even the second-order PN effects are being investigated \citep{deng2015second,deng2016second}.
The PPN formalism is a promising approach in testing cosmological models ~\citep{sanghai2015post}.
\par
{
Post-Newtonian equations of motion also can be used for a possible solution of the well-known problem of pseudo-tensor of the energy-momentum tensor of the gravitational field in GR (\citet{landau1971} (Par. 101 "The energy-momentum pseudotensor", p.304)
and review \citet{baryshev2017}). The point is that in the frame of the Feynman's relativistic quantum-field approach to gravitational interaction ~\citep{Feynman1995} there is ordinary
(not "pseudo") energy-momentum tensor $T^{\mu\nu}$ of the gravity field. So there is positive localizable energy density of the gravitational field $T_g^{00}$, which in the case of the PN approximation equals
$(\bm{\nabla}\varphi_N)^2/
8\pi G \,\, \,\,(erg/ cm^{3})$.
Fortunately, this energy-mass distribution around central massive body gives 16.7\% contribution to the relativistic pericenter shift and together with linear part gives the standard value of the Mercury perihelion shift~\cite{baryshev2002}.
}


The S-stars were being used in a big amount of different investigations. The most interesting in the context of this work are investigations of the motion of the S-stars. The possibility of detecting the Post-Newtonian effects in the motion of S2 was being researched some time ago \citep{preto2009post}. It was shown that the perturbations produced by these effects are significant and have a specific time dependence. In the work \citet{parsa2017investigating} the authors used the PN approximation to calcuate the stellar orbits and investigate how do they change if the fitting is employed in different parts of the orbit. The important effects that can be seen in the S-stars data are the effects related to the light propagation. It is possible to investigate the relativistic redshift using the spectroscopic data. The authors of work \citet{saida2019significant} show that there is a double-peak-appearance in the time evolution of the defined GR evidence.
\par
The purpose of this work is to consider the relativistic Post-Newtonian motion of several S-stars, using the data from the work \cite{Me2020},
{ and
to attract attention to the theoretical uncertainty in analysis of PN motion of S-stars due to coordinate dependence of the orbital shapes of the PN test body motion
in GR. 
Also in the frame of the Feynman's field-theoretical approach (FGT) we estimate the energy density of the static gravitational field for S-stars conditions.}  

{ The effects related to spin of the central massive object, such as Lense-Thirring precession, are also detectable via observations of the most recently discovered S-stars \citep{Peissker2020,Iorio2020,Fragione2020}. It is important that we consider only the Schwarzschild problem and do not take the spin-related effects into consideration, which we shall study in separate paper.}

\section{Observational data}

The S-stars cluster is being observed for 25 years now. This period of time is comparable with the orbital periods of the stars. Some of them have made a revolution of their orbit, and in particular cases it is even not a single time. The accumulated arrays of data differ in their meaningfullness for different S-stars. The star named S2 is the most popular choice among the researchers, and it is not unreasonable. The S2 is the most thoroughly observed, so it has the biggest time series of the observational data. We will consider this star, along with the stars S38 and S55.\par
There are two types of S-stars observational data: the astrometric and the spectroscopic one. The astrometric data is given as the visual positions of the S-stars on the sky. It is important that one should take into account not only the pair of coordinates, but also the time of the observations. At first sight it {may} seem {to the reader} that { to obtain the best-fit trajectory, one should implement} the fitting procedures { that} use the points of visual positions of the stars to fit the ellipses into them as the visual trajectories. But what do these procedures actually fit are not { 2D} ellipses, but spirals { in the 3D space with the axes of not only right ascension and declination, but also the time of the observed positions.} The spectroscopic data consists of radial velocities, corrected with respect to the Local Standard of the Rest (the so-called VLSR-correction) of the S-stars. But we cannot use the modelled velocities directly to fit the observed radial velocities. We shall interpret the latter as the redshift values. The modelled radial velocities must be transformed to observable values with the equation (\ref{eq:RVtransform}) so the relativistic Doppler effect and the gravitational redshift are taken into account.\par
The dataset that we use is described in previous work \cite{Me2020}. It is a dataset combined from the data of 4 works \citet{Gillessen_2017}, \citet{Boehle_2016}, \citet{Chu_2018}, and \citet{Do664}. It contains the astrometric and the spectroscopic data of S2, S38 and S55 obtained on VLT, Keck observatory and Subaru telescope. The data presented in different works must be corrected for difference in the reference system \citep{Gillessen_2017,GRAVITY2020}.

\section{Post-Newtonian effects}
\subsection{Equations of motion}
For the purpose of this work it is important to use the equations of motion, which will be integrated to simulate the S-stars trajectories. In the frame of geometrical General Relativity theory (GR), according to classical textbook by \citeauthor{Brumberg1991} (1991) , the Post-Newtonian equations of motion of a test particle in the static gravitational field of central massive body are given by
\begin{equation}
    \label{eq:PNBrum}
    \ddot{\mathbf{x}} =-\bm{\nabla}\varphi_N \bigg( 1 + (4-2\alpha)\frac{\varphi_N}{c^2} + (1+\alpha)\frac{\dot{\mathbf{x}}^2}{c^2} - 3\alpha\frac{(\mathbf{x}\cdot\dot{\mathbf{x}})^2}{c^2x^2} \bigg) + (4-2\alpha)\bigg( \bm{\nabla}\varphi_N \cdot\frac{\dot{\mathbf{x}}}{c} \bigg)\frac{\dot{\mathbf{x}}}{c},
\end{equation}
where $\alpha$ denotes the coordinate system choice. For example, $\alpha=1$ for the Standard (Schwarzschild) coordinates 
{
and for either isotropic or harmonic coordinates the parameter $\alpha=0$. 
Note that, though the secular effects do not depend on the parameter $\alpha$, the shape of the orbit is different for different $\alpha$.
For the purposes of the relativistic celestial mechanics, the isotropic coordinate system is more commonly used, as it is stated in \citet{MTW1973}. 
Here it is important to note that modern S-stars observations allow, at first time, to measure directly the shape of the S-star orbit, which means that to test the value of the the parameter $\alpha$ itself. This rises a question about the role of the coordinate system choice in GR.
}

For isotropic or harmonic coordinates the parameter $\alpha=0$, so the corresponding PN equations of motion in the frame of GR are
\begin{equation}
    \label{eq:EqMotPN}
    \ddot{\mathbf{x}} = -\bm{\nabla}\varphi_\textrm{N} \bigg( 1 + 4\frac{\varphi_\textrm{N}}{c^2} + \frac{\dot{\mathbf{x}}^2}{c^2} \bigg) + 4 \Big( \bm{\nabla}\varphi_\textrm{N}\cdot \frac{\dot{\mathbf{x}}}{c} \Big)\frac{\dot{\mathbf{x}}}{c}.
\end{equation}
{
Intriguingly, these equations of motion coincide with the PN equations of motion in the Feynman's field-theoretical approach to gravitation (FGT) (\citep{Baryshev1986,baryshev2002}), where they do not depend on the choice of coordinate system. Importantly, the term 
$4\frac{\varphi_\textrm{N}}{c^2}$ includes the  non-linear part: factor $(4 = 3+1)$, where 3 is the linear part and 1 is the non-linear part. This non-linear part corresponds to the contribution of the effective radial mass-energy density distribution of the gravitational field itself around the central body
\begin{equation}
\label{grav-energy-density}
T_g^{00}=\epsilon_g=(\bm{\nabla}\varphi_N)^2/ 8\pi G \,\, \,\,(erg/ cm^{3}).    
\end{equation}
This additional mass distribution gives correction to the 00-component of the tensor gravitational potential 
$\psi^{\mu\nu}$
in the form:
\begin{equation}
\label{phi-corr}
\psi^{00}=\phi= \varphi_N +
\frac{1}{2}
\frac{\varphi^2_\textrm{N}}{c^2}\,\,,
\end{equation}
which leads to the non-linear part of the relativistic pericenter shift.
Derivations of all needed equations and possible applications of the PN equations of motion are presented in review
\cite{baryshev2017}.
}

The PPN generalization of the Schwarzschild metric in the frame of isotropic coordinates leads to the PPN equations of motion \cite{Me2020}, which are given by
\begin{equation}
    \label{eq:EqMotPPN}
    \ddot{\mathbf{x}} = -\bm{\nabla}\varphi_\textrm{N} \bigg( 1 + 2(\beta+\gamma)\frac{\varphi_\textrm{N}}{c^2} + \gamma \frac{\dot{\mathbf{x}}^2}{c^2} \bigg) + 2(\gamma+1) \Big( \bm{\nabla}\varphi_\textrm{N}\cdot \frac{\dot{\mathbf{x}}}{c} \Big)\frac{\dot{\mathbf{x}}}{c}.
\end{equation}
Substituting $\beta = \gamma = 1$ leads to GR PN equations in the frame of the isotropic or harmonic coordinate systems. 
We can see that in the limit $c\rightarrow\infty$ these equations reduce to Newtonian equations of motion
\begin{equation}
    \label{eq:EqMotN}
    \ddot{\mathbf{x}} = -\bm{\nabla}\varphi_\textrm{N}.
\end{equation}
In the work of \citet{Brumberg1991}, the author considered the problem in a more general way. He did also obtain PPN equations of motion, but the coordinate system that he considered was not fixed. This fact implies that there appears the corresponding third degree of freedom, which is related to the coordinate system choice. So \citeauthor{Brumberg1991} used to have 3 PPN parameters $(A, B, K)$, instead of the 2 usual PPN parameters $(\beta, \gamma)$.

\subsection{Pericenter shift}
The Post-Newtonian equations of motion lead us into the effect of the pericenter shift. The value of the pericenter shift per one turn in GR is well known \cite{Brumberg1991}, \cite{Damour1985}
\begin{equation}
    \label{eq:pershift}
        \Delta\omega = \frac{6\pi r_\mathrm{g}}{a(1-e^2)},
\end{equation}
where $a$ is a semi-major axis, $e$ is an eccentricity and $r_\mathrm{g} = {GM}/{c^2}$ is a gravitational radius.\par
From coincidence of the PN equations of a test particle motion in GR and FGT it follows that the expression for the pericenter shift in FGT and GR is exactly the same. The difference from GR is that the FGT formula has two different terms:
\begin{equation}
    \label{eq:pershift_FGT}
        \Delta\omega = \frac{6\pi r_g}{a(1-e^2)} = \frac{7\pi r_g}{a(1-e^2)} - \frac{\pi r_g}{a(1-e^2)}
\end{equation}
The first term with a factor of $7\pi$ corresponds to the linear approximation, when in the field equations one does not take into account the non-linear contribution (energy of gravitational field itself). The second term with a factor of $\pi$ occurs after taking into account the non-linearity due to the positive energy density of the gravitational field. This term corresponds to the measurement of the field energy of the gravitation via pericenter shift observations.\par
The pericenter shift values and parameter of $\frac{\varphi_N}{c^2}$ in pericenter predicted for S-stars are:
    \begin{table}[H]
        \centering
        \begin{tabular}{lccc}
            \hline
            Star            & S2 & S38 & S55 \\ \hline
            $\Delta\omega$  & $12'$ & $7.1'$ & $6.7'$ \\
            $\dot{\omega}$  & $45''\textrm{/yr}$ & $22''\textrm{/yr}$ & $31''\textrm{/yr}$ \\
            $\dot{\omega}\cdot100\;\textrm{yrs}$ & $1^\circ15'$ & $37'$ & $52'$ \\
            $\big(\frac{\varphi_N}{c^2}\big)_{\textrm{per}}$ & $-3.48\cdot10^{-4}$ & $-1.99\cdot10^{-4}$ & $-1.69\cdot10^{-4}$ \\ \hline
        \end{tabular}
        \caption{S-stars pericenter shift}\label{tab:Sstars-pericenter}
    \end{table}
 \noindent The observable effect is similar to one with \textbf{Mercury anomalous pericenter shift}, that Einstein explained. { The effect of GR Schwarzschild perihelion precession was also measured with artificial Earth satellites \citep{Lammerzahl2004,Lucchesi2010}, so this effect is well studied in the Solar System. } Now we can observe it with S-star cluster in the center of our Galaxy.
For Mercury, the pericentral $\frac{\varphi_N}{c^2}$ is $-3.21\cdot10^{-8}$, and for binary pulsar PSR 1913+16 it is $-2.7\cdot10^{-6}$. We can see that the field at the pericenter of the S-stars orbits is stronger by 2 orders of magnitude.\par
    Although the values of the pericenter shift seem to be small, they are actually detectable{, which is shown in the recent work of} \citet{GRAVITY2020}. The pericenter shift appears as a consequence of the PN equations of motion. That means that using the PN equations of motion to fit the trajectories implies detecting the pericenter shift effect. In the frame of field approach it means that we are able to measure the field energy density, which is positive and localizable.

\subsection{Light Propagation}
We must take into consideration several relativistic effects related to light propagation. Such as the relativistic Doppler effect, which is the reason we must transform modelled RV’s to the observable RV’s. { Another significant light propagation effect is }the gravitational redshift{, which was successfully measured by \citet{Do664}}. Taking both of these effects into account is described in \cite{Me2020} by the equation
\begin{equation}
    \label{eq:RVtransform}
    \mathrm{RV}_\textrm{obs}/c = \mathrm{RV}_\textrm{model}/c + \frac{\varphi_\textrm{N}}{c^2} + \frac{v^2}{2c^2}.
\end{equation}
{The detailed study of this effect is described in the work of \citet{Iorio2017}.}\par
It is also necessary to take into account the R\o mer delay. The gravitational lensing and Shapiro time delay effects are negligible in this problem.
\section{Modelling}
\subsection{Parameters}
Our aim is to build a model of the S-stars observations which depends on PPN parameters $\beta$ and $\gamma$. After that, we will be able to use { some of the optimization} techniques to solve the inverse problem and estimate the parameters. So the question is how many parameters do we need to build a model.\par
To define the orbit and the position of the star, we need 6 parameters. For example, these could be Keplerian parameters or the components of the initial phase vector. { The coordinate system that we use is defined as follows: the $xy$ plane is the sky itself, where $y$ axis points in the direction of declination, $x$ axis has the reversed direction of right ascension, and the $z$ axis is pointed from Sgr A* to the observer.} The next 2 parameters that we need are the mass $\mathfrak{M}$ of Sgr A* and the distance $R_0$ to the Galactic Center. The Sgr A* itself also has the velocity relative to our Solar System, so we have two more parameters of its initial position on the sky {(which alongside with the distance $R_0$ form the vector of the Sgr A* position relative to the Solar System.)} and 3 parameters, which define its velocity. The dataset that we use contains 3 different {observational data} arrays, that { were obtained by different teams, and hence} have their own different reference frames. { The biggest observational dataset is the one from \citet{Gillessen_2017}. We will use it as the reference observational dataset aligned with our coordinate system. As we have two more datasets (from the papers \citep{Boehle_2016,Do664}), we must add 4 more parameters, which are the RA and Dec offsets from the dataset of \citep{Boehle_2016} or \citep{Do664} and the reference dataset of \citep{Gillessen_2017}.} And the final 2 of our parameters are the PPN parameters of $\beta$ and $\gamma$ that we want to constrain. In total we have 19 parameters if we consider 1 star, or 31 parameters if we consider 3 stars. { It is a large amount of parameters. In the next section it is shown that the model construction implies very non-trivial dependency of the likelihood function from the parameters. Because of this along with the large amount of parameters, we decided to use the bayesian inference techiques, such as the MCMC sampling.}

\subsection{Constructing a model}
The first step in the modelling is to find the trajectory. We want to integrate the equations of motion in a 2D plane. For this purpose, we need 4 parameters of initial phase vector components $(x_0,y_0,\dot{x}_0,\dot{y}_0)$. They are used as the initial conditions for the integrator, which is the Owren-Zennano 4/3-order method implemented in the DifferentialEquations package {\cite{rackauckas2017differentialequations}} of the Julia language {\cite{bezanson2017julia}}. It produces the array of phase vectors $(x,y,\dot{x},\dot{y})$, which is a modelled trajectory. To rotate it, we need 2 angles: the longitude of the ascending node $\Omega$ and the inclination $i$, which are the remaining 2 parameters. {To perform the rotations, one should pad the modelled array of phase vectors $(x,y,\dot{x},\dot{y})$ to 6D phase space $(x,y,z,\dot{x},\dot{y},\dot{z})$, where $z$ and $\dot{z}$ are zeros. After that one should multiply each triplet of coordinates $(x,y,z)$ and velocities $(\dot{x},\dot{y},\dot{z})$ with the rotation matrices that contain $\Omega$ and $i$. As the result, one will obtain the modelled array of phase vectors in the reference frame that is needed.} In total we have 6 parameters that define the orbit and the position of the star, as it was stated before.\par
The next step is getting RA and Dec from $x$ and $y$ of modelled trajectory. The R\o mer time delay is taken into account at this step. We shall divide the coordinates by $R_0$, but one important thing is the sign. We should inverse $x$, as RA scale is clockwise. And we also must take into account the proper motion of Sgr A*.\par
And finally, we have to obtain RV from $\dot{z}$, using the equation (\ref{eq:RVtransform}) due to the light propagation effects.

\section{MCMC analysis}
{Bayesian inference is the statistical method of inference in which the probability of a hypothesis is updated with the Bayes's theorem
\begin{equation}
    \label{eq:bayes}
    P(H|E) = \frac{P(E|H)P(H)}{P(E)},
\end{equation}
where $H$ stands for the hypothesis and $E$ stands for evidence. The $P(E|H)$ is a likelihood. It is a probability of observing the data (evidence) given hypothesis. $P(E)=P(E|H)P(H)+P(E|\neg H)P(\neg H)$ is called model evidence or the marginalised likelihood. $P(H)$ stands for the prior probability (the probability of hypothesis before taking the observed data into the account) and $P(H|E)$ is the posterior probability (after taking the data into account). In the context of the bayesian parameter inference, the Bayes's theorem may be interpreted as
\begin{equation}
    \label{eq:bayespar}
    P(\bm{\theta}|\mathbf{X}) = \frac{P(\mathbf{X}|\bm{\theta})P(\bm{\theta})}{P(\mathbf{X})} \propto P(\mathbf{X}|\bm{\theta})P(\bm{\theta}),
\end{equation}
where $\mathbf{X}$ is a sample of the observed data and $\bm{\theta}$ is a vector of the parameters. $P(\bm{\theta})$ is a prior distribution of the parameters, $P(\mathbf{X}|\bm{\theta})$ is the likelihood and $P(\bm{\theta}|\mathbf{X})$ is a posterior distribution of the parameters.\par}
Once the model is defined, it is possible to simulate the data values { and compare them with the real observable values. To do} that, one can define the chi-square residuals and the likelihood function. The latter { can be} used by the Markov Chain Monte Carlo sampler. { We use the one} implemented in the {AffineInvariantMCMC} package {(documentation available at https://github.com/madsjulia/AffineInvariantMCMC.jl)} for The Julia language {\cite{bezanson2017julia}}. Using the MCMC sampling, one can obtain the constraints of parameters of a defined model.\par
In the previous work \cite{Me2020}, some of the parameters were considered to be fixed, so the model was simplified and had the less amount of the parameters. However, in the work~\citep{GRAVITY2020} it is stated, that this simplification leads to the uncertainties. In our case, this is taken into account and the model now depends on 31 parameters. The process was launched with 300 { independently computed chains (the so-called 'walkers'), where each chain consists of the parameters and corresponding log-likelihood values for 2,000 iterations.}
\begin{figure}[H]
\centering
\includegraphics[width=10 cm]{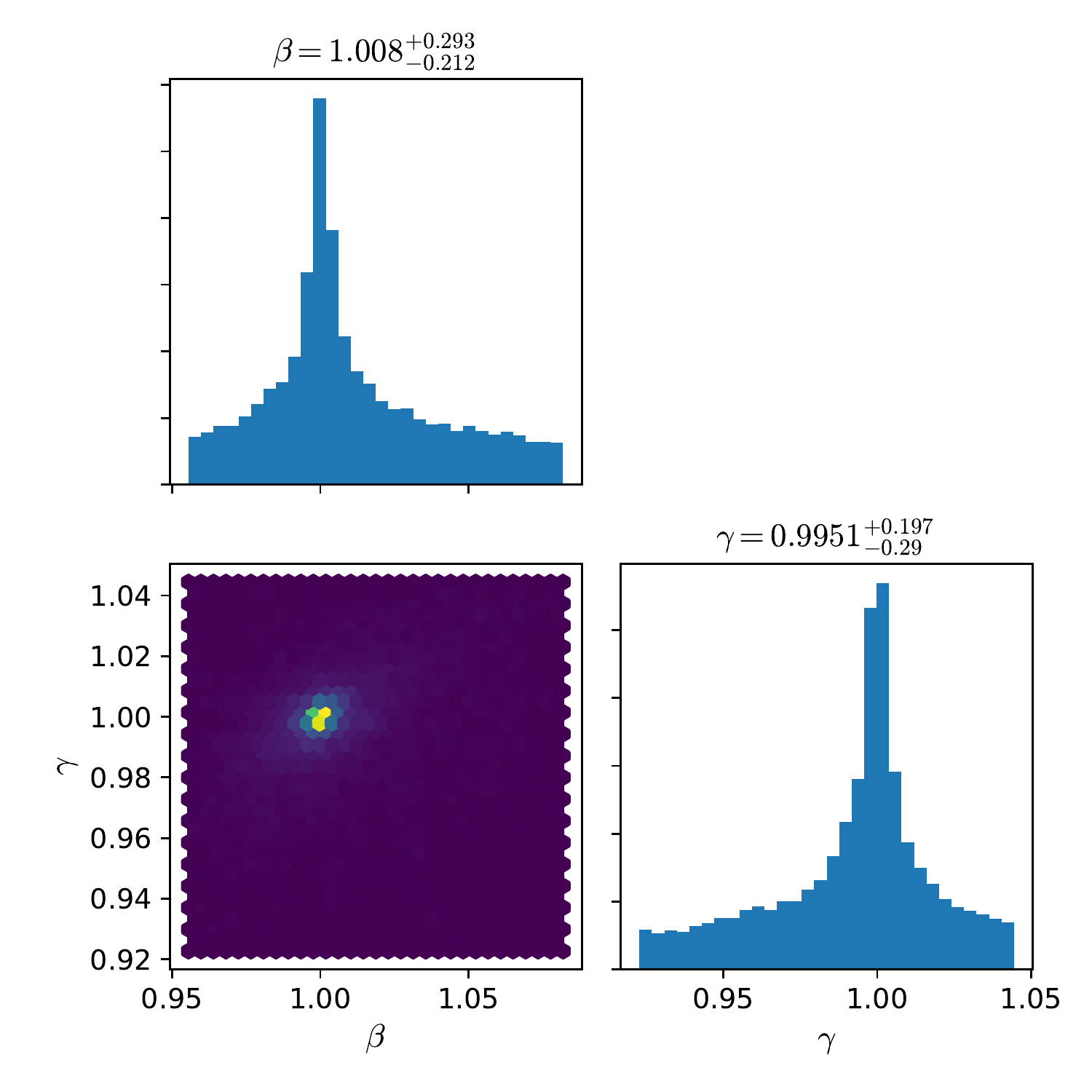}
\caption{The posterior distribution of $\beta_\textrm{PPN}$ and $\gamma_\textrm{PPN}$.}
\label{fig:PPNpars}
\end{figure}
The posterior distributions of the $\beta_\textrm{PPN}$ and $\gamma_\textrm{PPN}$ are presented in the Figure \ref{fig:PPNpars}. { The right and top plots show us the histograms of the posterior distributions of the considered parameters. The bottom-left plot is a '2D-histogram' of this parameters, where the color shows us the amount of points in a bin. This '2D-histogram' is actually a projection of our 31-dimensional posterior distribution of the parameters to the 2D plane of $\beta_\textrm{PPN}$ and $\gamma_\textrm{PPN}$ parameters.} We can see that the PPN parameters don't make a significant contribution to the model. So these parameters have relatively high errors. But it is possible to constrain the PPN parameters with this approach. The errors are much smaller than in the previous work \cite{Me2020}.
\begin{figure}[H]
\centering
\includegraphics[width=10 cm]{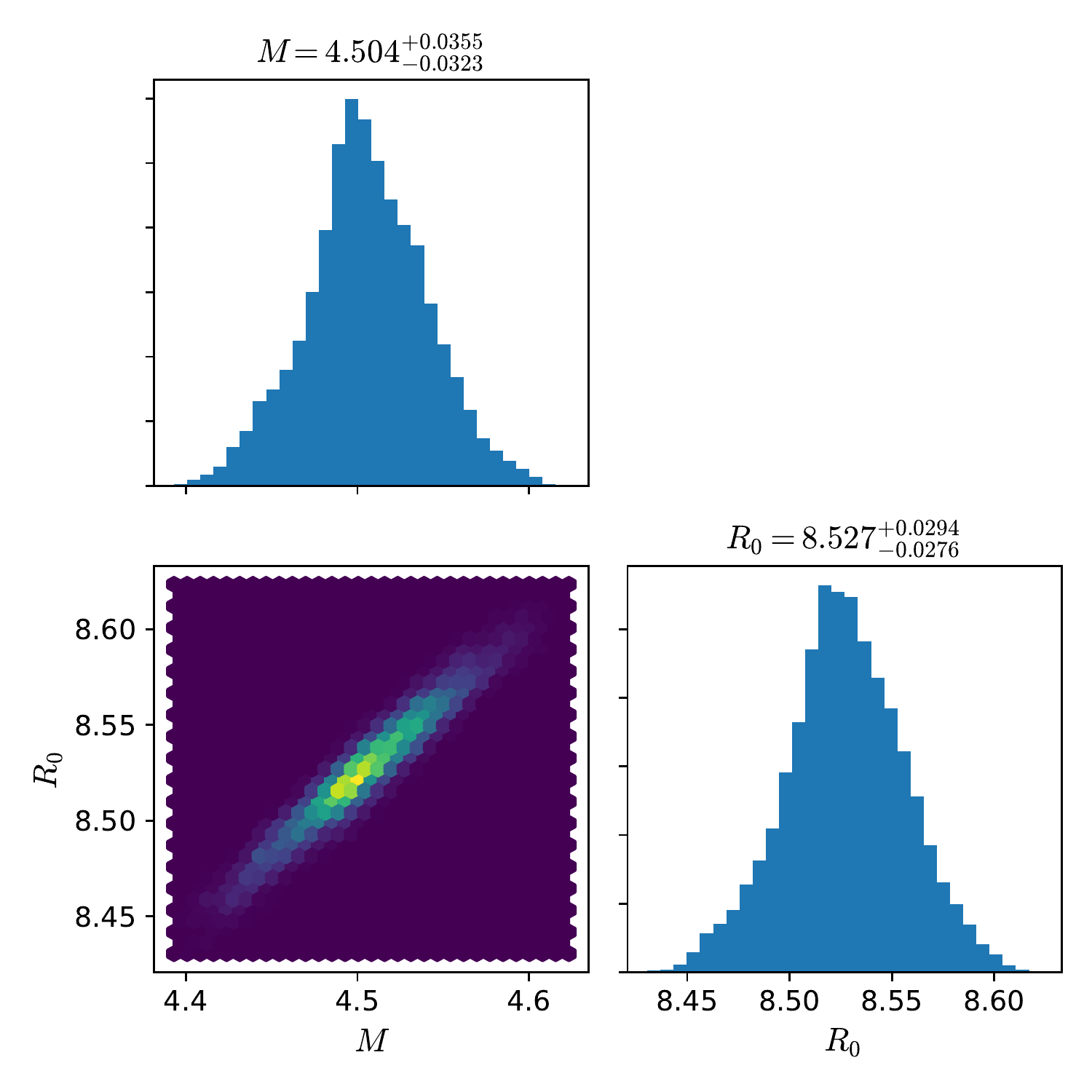}
\caption{The posterior distribution of $\beta_\textrm{PPN}$ and $\gamma_\textrm{PPN}$.}
\label{fig:MR}
\end{figure}
The other important parameters constrained by our model are the Mass of Sgr A* { in the units of $10^6\,M_\odot$} and the distance to the Galactic Center { in kpc}. The posterior distributions of these parameters are presented in the figure \ref{fig:MR}. { In this posterior distribution it is clearly visible that there exists a correlation between these parameters, as the posterior distribution points line up.}
Orbital parameters of S2, S38, S55; and reference frame offset values are given in the Appendix \ref{app1}.
\begin{figure}[H]
\centering
\includegraphics[width=13 cm, clip]{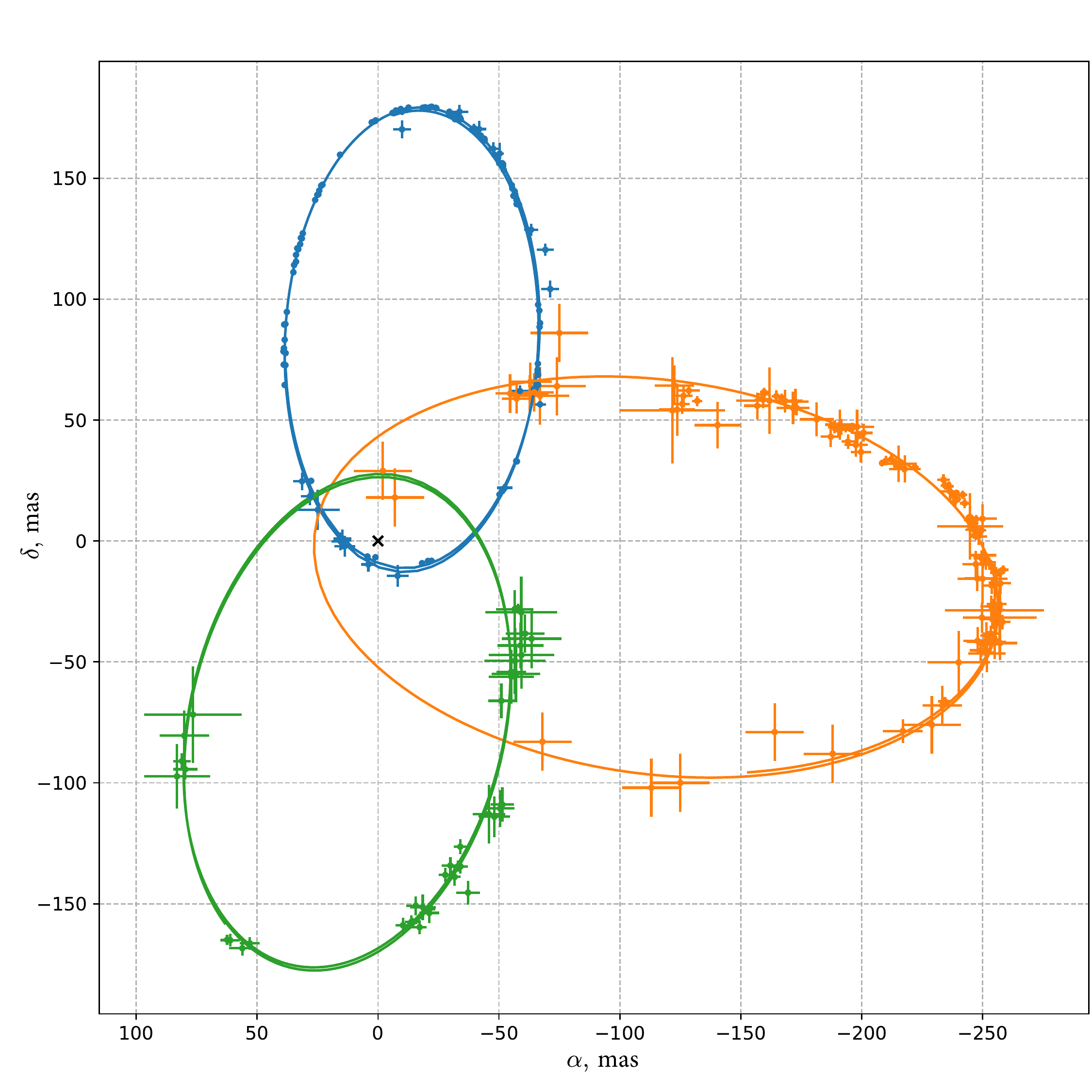}
\caption{The best-fit PPN orbits of S2 (blue), S38 (orange) and S55 (green).}
\label{fig:traj}
\end{figure}
\begin{figure}[H]
\centering
\includegraphics[width=10 cm]{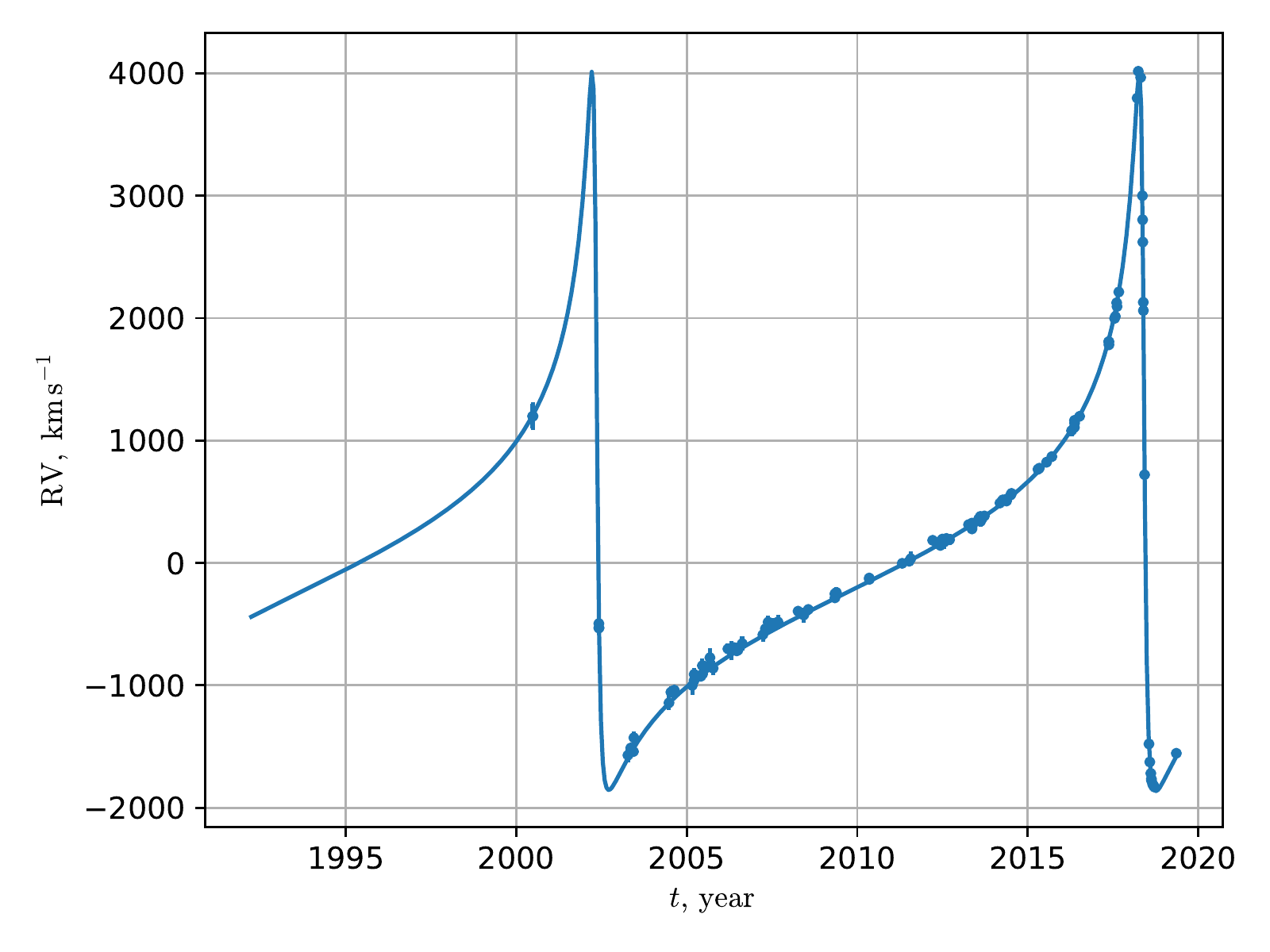}
\caption{The RV plot of best-fit PPN orbit of S2.}
\label{fig:RV}
\end{figure}
\begin{figure}[H]
\centering
\includegraphics[width=10 cm]{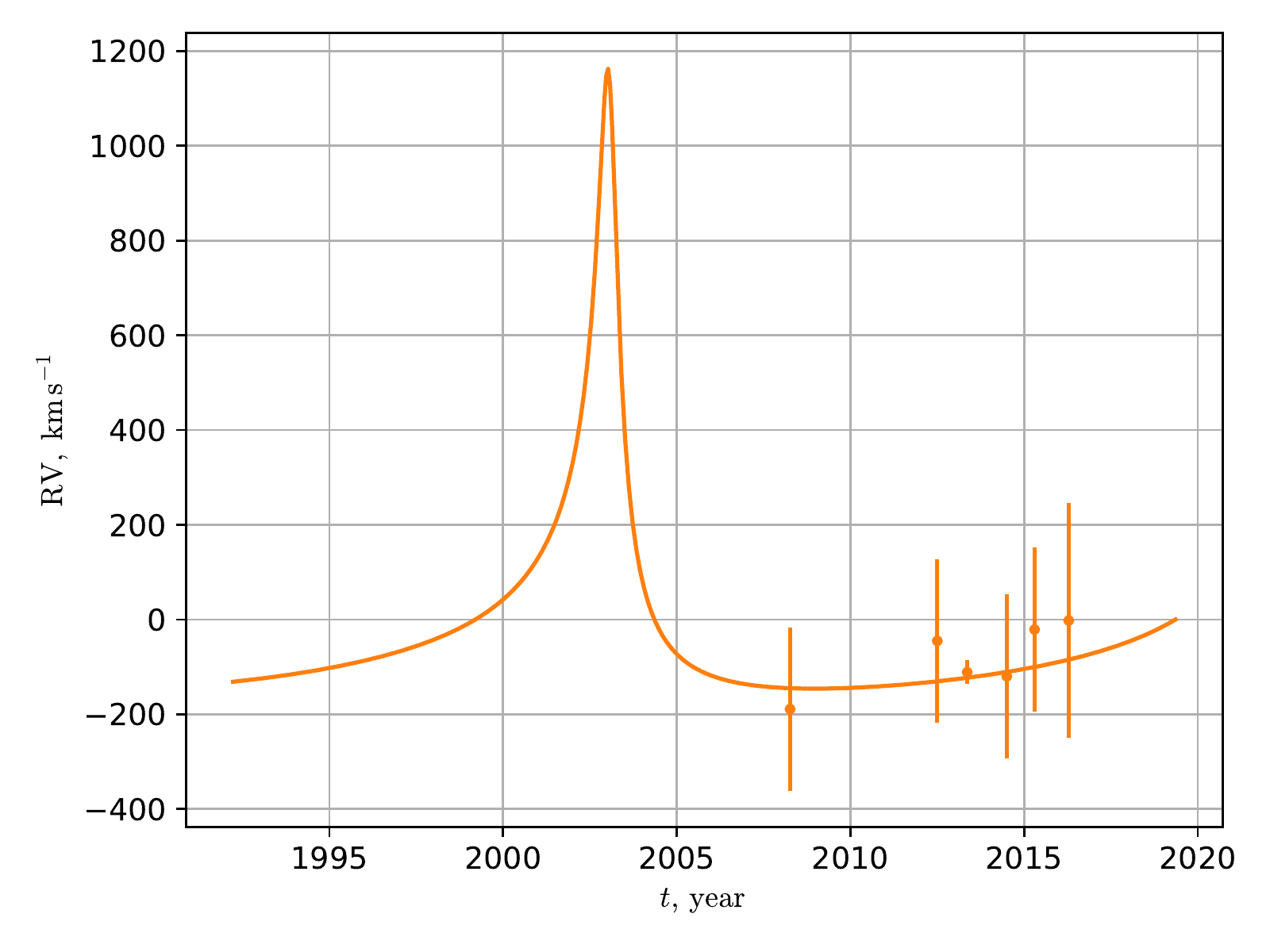}
\caption{The RV plot of best-fit PPN orbit of S38.}
\label{fig:RV2}
\end{figure}
\begin{figure}[H]
\centering
\includegraphics[width=10 cm]{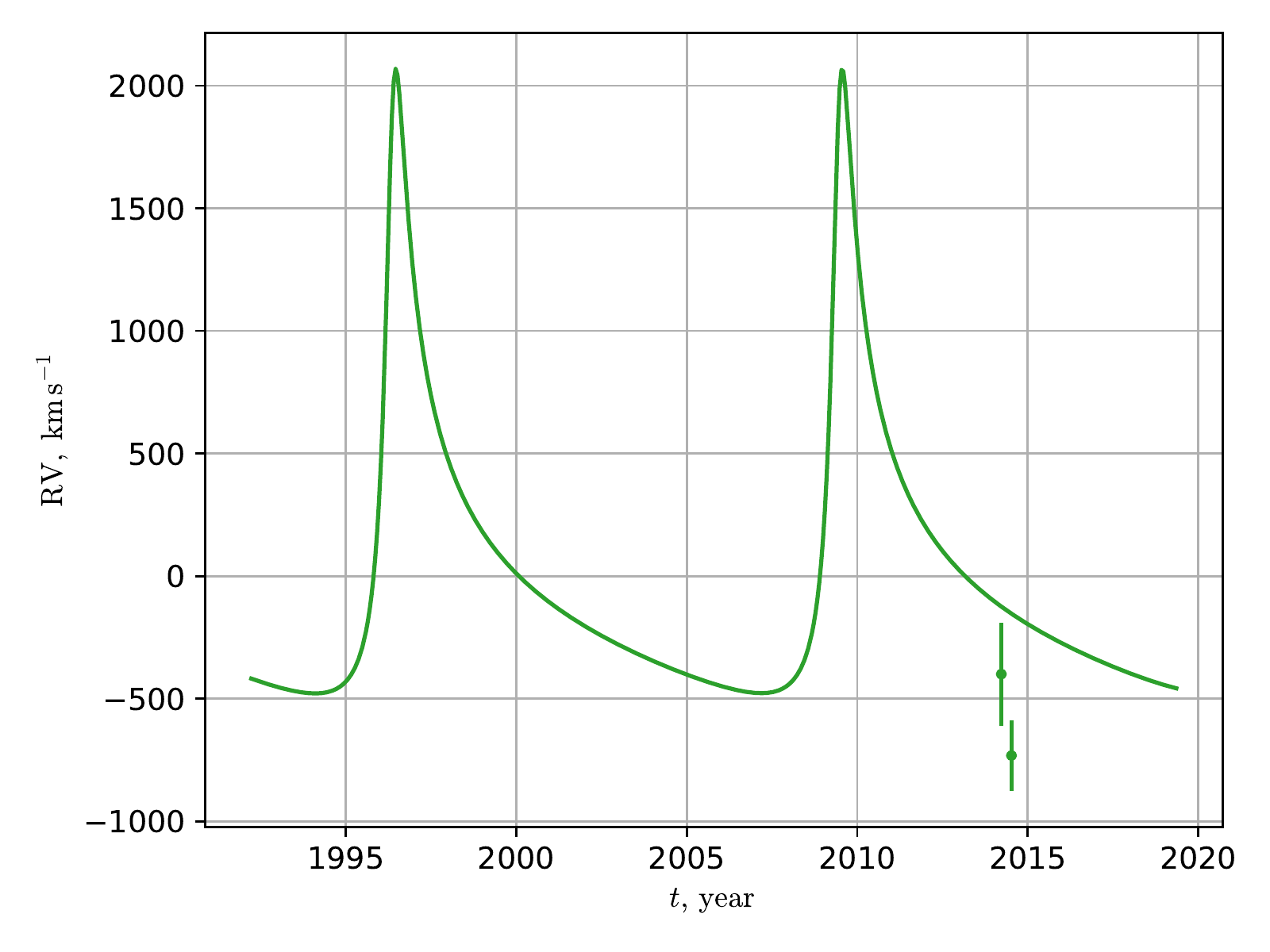}
\caption{The RV plot of best-fit PPN orbit of S55.}
\label{fig:RV3}
\end{figure}
The Figures \ref{fig:traj}, \ref{fig:RV}, \ref{fig:RV2} and \ref{fig:RV3} show us the best-fit orbit trajectories and RV plots. We can see that the trajectories are { non-closed}. That is not caused by the pericenter shift, but by the proper motion {(transverse velocity)} of the Galactic Center relative to Solar System.

\section{Tidal disruption}
Such a strong gravitational field also produces significant tidal forces. The question is how strong are these forces in relation to the objects orbiting the Sgr A*. We can calculate the Roche limit { for the fluid satellite (the distance of its tidal disruption) near} the Sgr A*.{
\begin{equation}
    \label{eq:Roche}
    d = 2.44R_*\bigg( \frac{M_\textrm{SgrA*}}{M_*} \bigg)^\frac{1}{3} \approx 2.44R_\odot\bigg( \frac{M_\textrm{SgrA*}}{M_\odot} \bigg)^\frac{1}{3} = 2\,\mathrm{AU},
\end{equation}}
{where $R_*$ and $M_*$ are radius and mass of the star. We assumed them equal to the mass and radius of the Sun just for the estimation.} For S2, S38 and S55 the pericenter is $\sim 200\,\mathrm{AU}$. The tidal disruption is extremely unlikely. It is possible only for extended objects, which are weakly gravitationally bound. There was a suggestion that  the observed object G2~\citep{Pfuhl2015G2} is a gas cloud.  But after the pericenter passage the G2 object conserved without tidal disruption, and it became clear that it is a star.

\section{Discussion \& Conclusion}
In this work the PPN parameters $\beta$ and $\gamma$ are constrained using the S-stars data. The difference from the previous work \cite{Me2020} is that the helpful remarks from \citet{GRAVITY2020} were taken into account. This time the parameters of ${\mathfrak M}$, $R_0$, etc. were not fixed, so they were constrained along with the other parameters. The dataset reference systems difference was also taken into account. The offsets between the reference frames were considered as the varying parameters for the MCMC sampling.\par
The estimates of PPN parameters in viscinity of the supermassive relativistic compact object Sgr A* are $\beta=1.01^{+0.29}_{-0.21}$ and $\gamma=0.99^{+0.19}_{-0.29}$. { The PPN scaling parameter of $f_\textrm{SP} = (2 + 2\gamma - \beta)/3$ is estimated to be $1.01^{+0.20}_{-0.17}$, which is comparable with the values obtained by \citet{GRAVITY2020}.} The result is consistent with the General Relativity predictions. The { formal estimation of} mass of Sgr A* { and the distance to the Galactic Center are} $\mathfrak{M}=4.50^{+0.04}_{-0.03} \cdot 10^6\,M_\odot$ and $R_0 = 8.53^{+0.03}_{-0.03}\,\mathrm{kpc}$.
{We note that the question about possible contribution of the choice of coordinate system in PN equations of motion needs further investigations.}

{
In the frame of the field-theoretical description of the gravitational interaction the observed pericenter shifts of the S-stars confirm the positive localizable energy density of the gravity field having value 
$(\bm{\nabla}\varphi_N)^2/
8\pi G $. Future studies of the motion of the most close to SgrA*  S-stars will test the gravity theories and give the crucial information on the nature of the central relativistic compact object SgrA*.}



\funding{This research received no external funding. The work was performed as part of the government contract of the SAO RAS approved by the Ministry of Science and Higher Education of the Russian Federation.}

\acknowledgments{Authors are grateful to S.~I. Shirokov 
for helpful advices and remarks.}
We thank the anonymous referee for their helpful suggestions. We also thank the GRAVITY collaboration for their useful comments on S2 data.


\abbreviations{The following abbreviations are used in this manuscript:\\

\noindent 
\begin{tabular}{@{}ll}
PPN     &   Parameterized Post-Newtonian \\
VLSR    &   Velocity of the Local Standard of Rest \\
MCMC    &   Markov Chain Monte Carlo \\
\end{tabular}}

\appendixtitles{no} 
\appendix
\section{}
\label{app1}
\begin{figure}[H]
\centering
\includegraphics[width=15.5 cm]{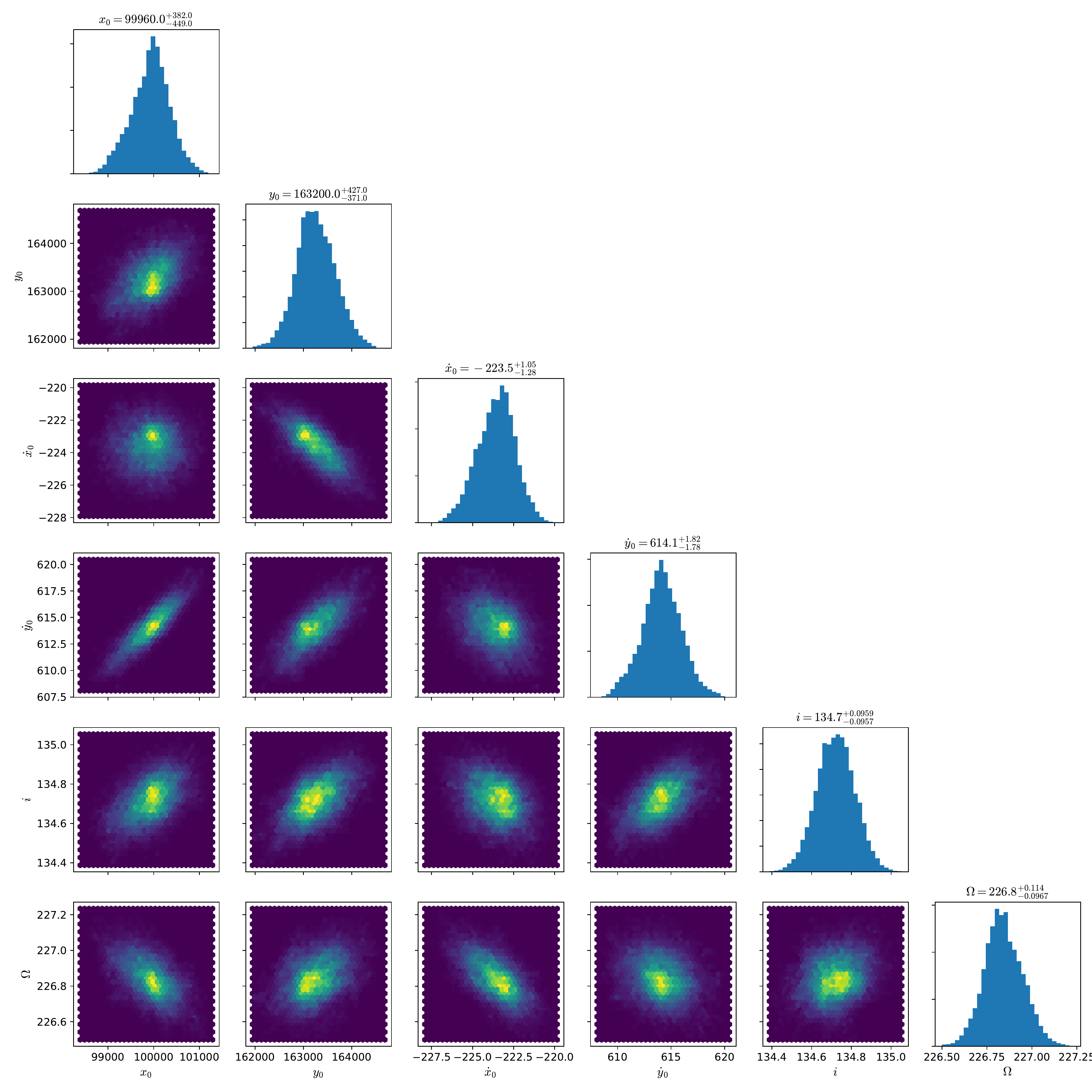}
\caption{The {posterior distribution of the} orbital parameters of the S2 star. { The initial position vector components $(x_0,y_0)$ are given in the units of the gravitational radius corresponding to the mass of $10^6\,M_\odot$ (which is $1476625\,\mathrm{km}$). The initial velocity components $(\dot{x}_0,\dot{y}_0)$ are given in $\mathrm{km}\,\mathrm{s}^{-1}$. The inclination $i$ and the longitude of the ascending node $\Omega$ are given in degrees.}}
\label{fig:S2orb}
\end{figure}

\begin{figure}[H]
\centering
\includegraphics[width=15.5 cm]{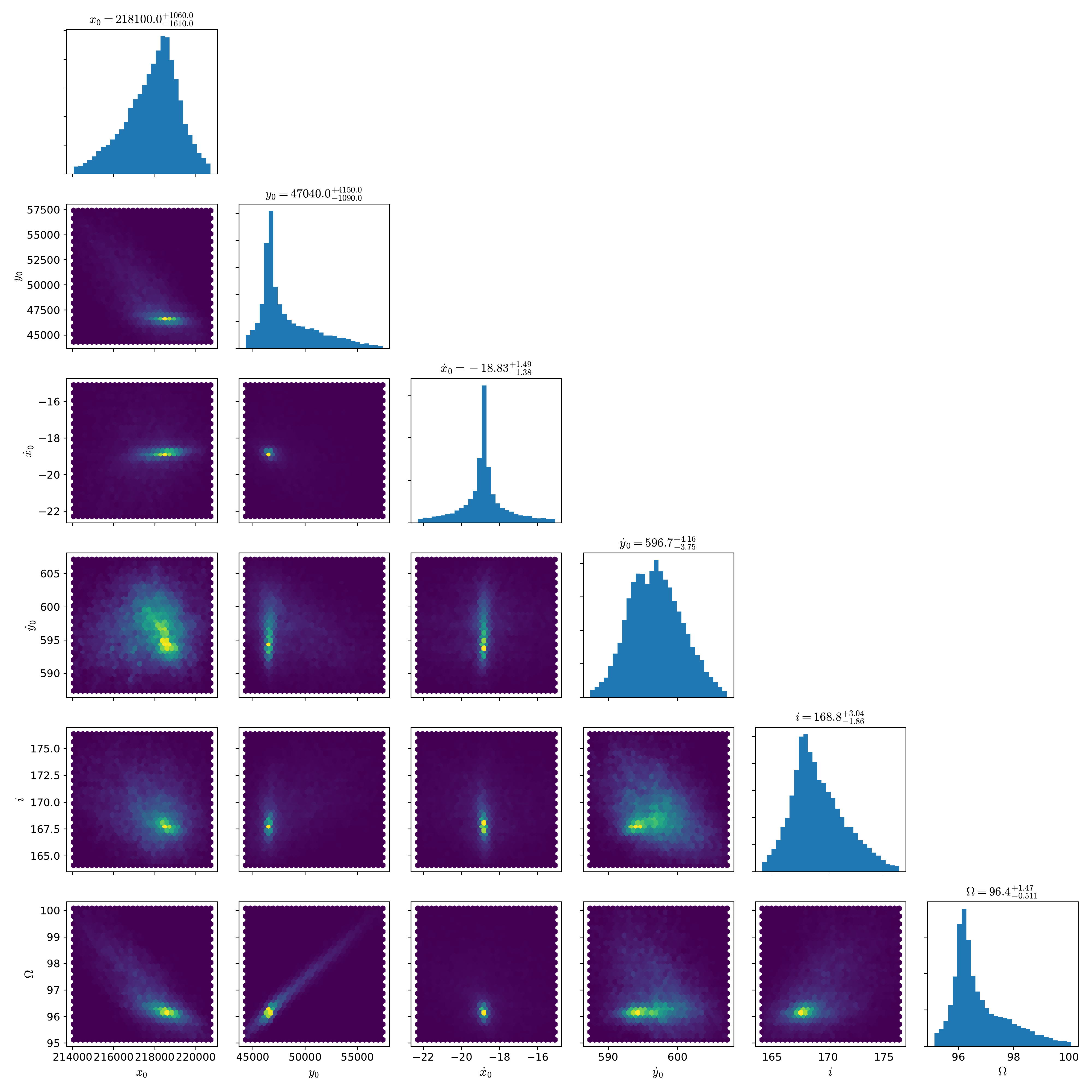}
\caption{The {posterior distribution of the} orbital parameters of the S38 star. {The units are the same as in Figure \ref{fig:S2orb}.}}
\label{fig:S38orb}
\end{figure}

\begin{figure}[H]
\centering
\includegraphics[width=15.5 cm]{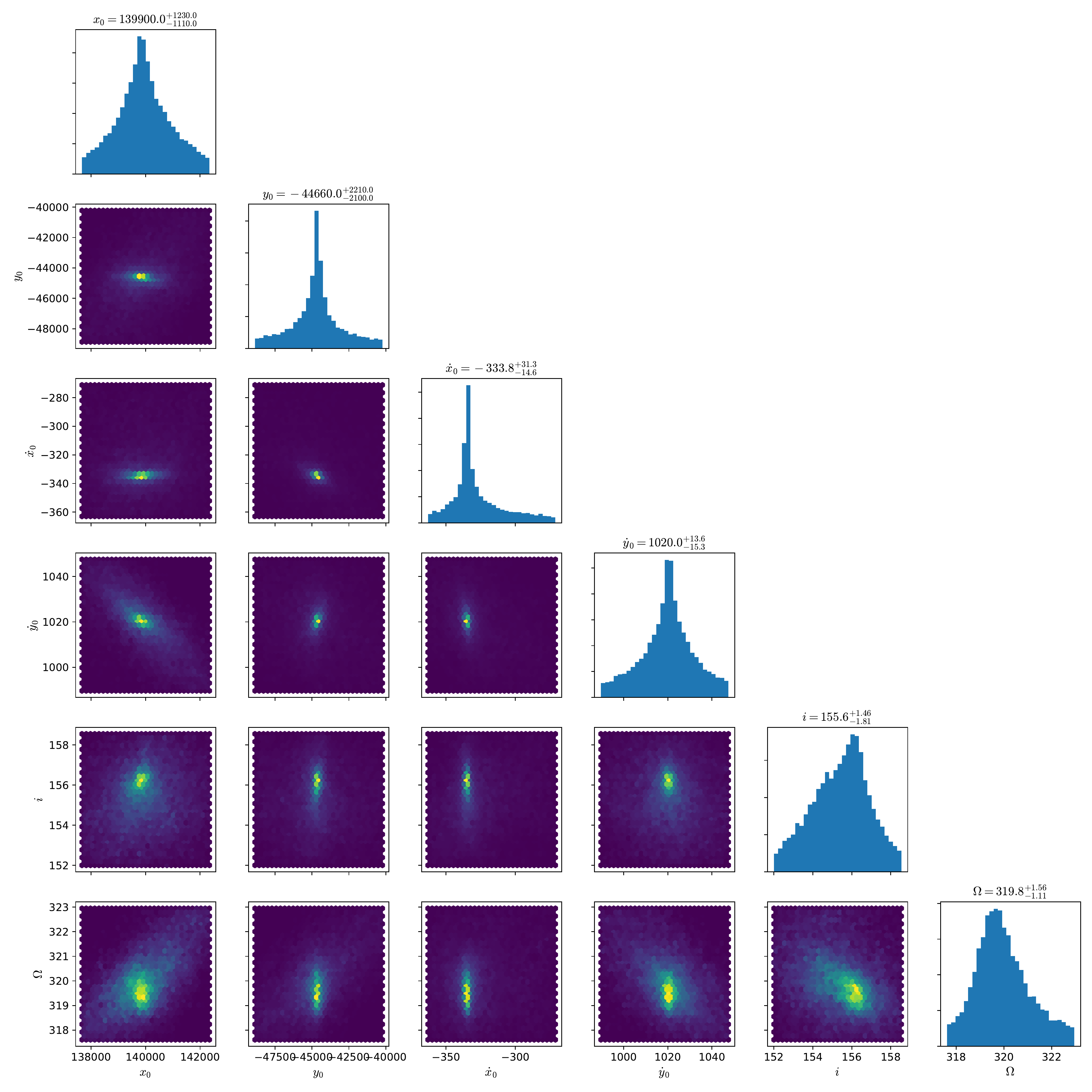}
\caption{The {posterior distribution of the} orbital parameters of the S55 star. {The units are the same as in Figure \ref{fig:S2orb}.}}
\label{fig:S55orb}
\end{figure}

\begin{figure}[H]
\centering
\includegraphics[width=15.5 cm]{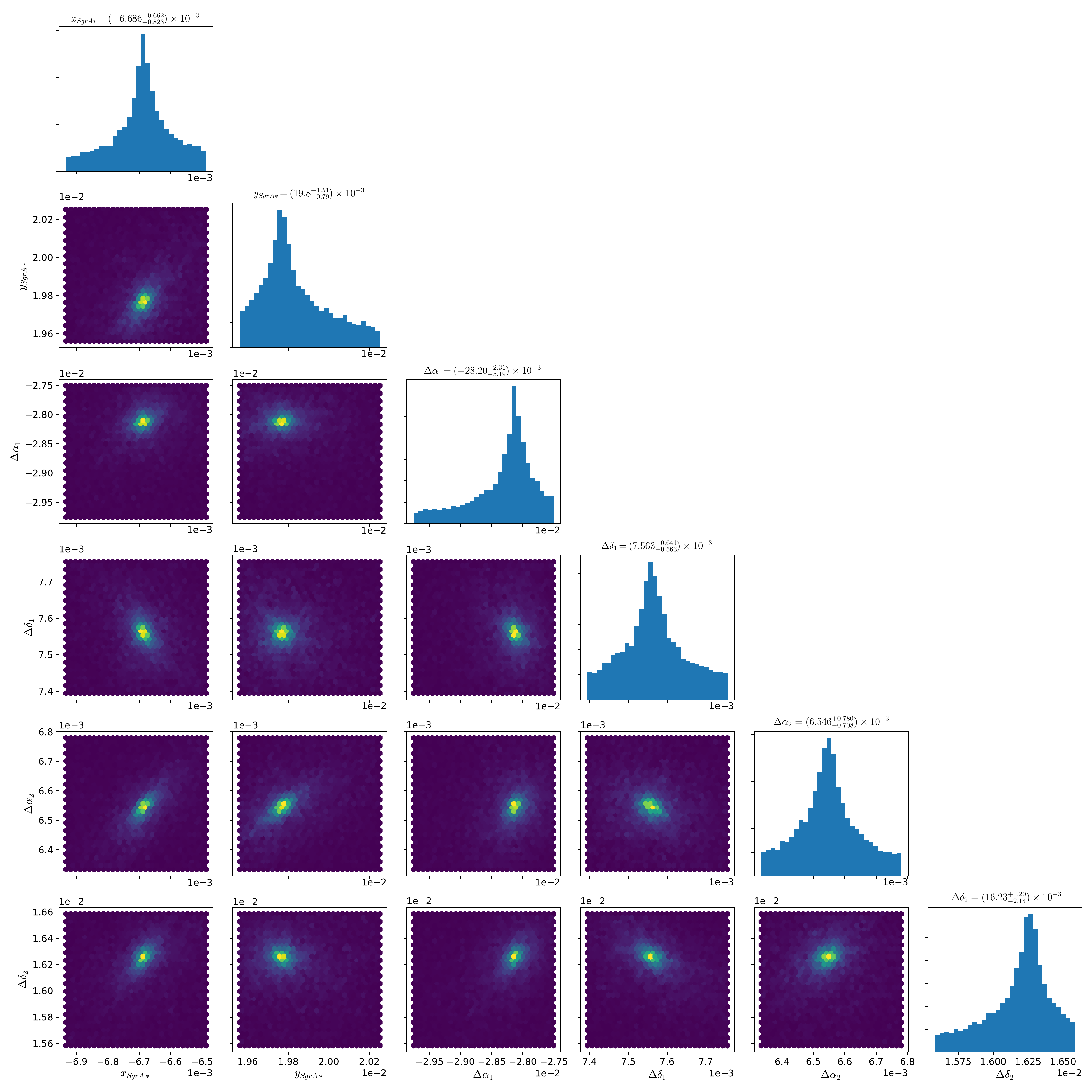}
\caption{The {posterior distribution of the} position of Sgr A* and reference frames offsets. {All values are given in the units of mas.}}
\label{fig:offsets}
\end{figure}

The posterior distributions of S2, S38, S55 and reference frames offsets are presented in the Figures \ref{fig:S2orb}, \ref{fig:S38orb}, \ref{fig:S55orb} and \ref{fig:offsets} respectively.



\reftitle{References}




\sampleavailability{Samples of the compounds ...... are available from the author.}


\end{document}